\documentclass[twocolumn,amsmath,amssymb,showkeys,showpacs]{revtex4}
\usepackage{amsmath}
\usepackage{amssymb}
\usepackage{subfigure}
\usepackage{graphicx}
\usepackage{dcolumn}
\usepackage{bm}
\usepackage{epsfig}
\usepackage{setspace}
 \usepackage{array}
\usepackage{braket}
\newenvironment{rcases}
  {\left.\begin{aligned}}
  {\end{aligned}\right\rbrace}

\newcommand{\bfg}{\begin{figure}}
\newcommand{\efg}{\end{figure}}
\begin{document}

\title {{\rm {\bf Multifragmentation model for the production of astrophysical strangelets}}}

\author {Sayan Biswas$^{a,b}$\footnote{Email : sayan@jcbose.ac.in}, J. N. De$^{c}$\footnote{Email : jn.de@saha.ac.in}, Partha S. Joarder$^{a,b}$\footnote{Email : partha@jcbose.ac.in}, Sibaji Raha$^{a,b}$\footnote{Email : sibaji.raha@jcbose.ac.in},
and Debapriyo Syam$^{b}$\footnote{Email : syam.debapriyo@gmail.com}}

\affiliation{$^{a}$ Department of Physics, Bose Institute, 93/1 A.P.C. Road, Kolkata, India 700009 \\
$^{b}$ Centre for Astroparticle Physics and Space Science, Bose Institute, Block EN, Sector V, Salt Lake, Kolkata, India 700091\\
$^{c}$ Saha Institute of Nuclear Physics, 1/AF, Bidhannagar, Kolkata 700064, India}

\begin{abstract}
\noindent Determination of baryon number (or mass) distribution of the strangelets, that may fragment out of the warm and excited strange quark matter ejected in the merger of strange stars in compact binary stellar systems in the Galaxy, is attempted here by using a statistical disassembly model. Finite mass of strange quarks is taken into account in the analysis. Resulting charges of the strangelets and the corresponding Coulomb corrections are included to get a plausible size distribution of those strangelets as they are produced in binary stellar mergers thus getting injected in the Galaxy. From this mass distribution of strangelets at their source, an approximate order of magnitude estimate for their possible flux in the solar neighborhood is attempted by using a simple diffusion model for their propagation in the Galaxy. Such theoretical estimate  is important in view of the ongoing efforts to detect galactic strangelets by recent satellite-borne experiments.

\end{abstract}


\maketitle
\section{Introduction}

\noindent Strange matter hypothesis (SMH)\cite{bib:witten} suggests that the strange quark matter (SQM), containing almost equal numbers of up ($u$-), down ($d$-) and strange ($s$-) quarks under spatial and color confinement within a phase boundary that separates the collection of quarks from the non-perturbative vacuum of quantum chromodynamics (QCD), may be the true ground state of hadronic matter. Finite size effects notwithstanding, small lumps of SQM (or strangelets) can also be more stable than ordinary nuclei~\cite{bib:farhi}. A possible scenario for the formation of strangelets in the Galaxy is the fragmentation of SQMs ejected in tidal disruptions of strange stars (SSs) in compact binary stellar systems~\cite{bib:mad2005}. Simulations of SS mergers~\cite{bib:bauswein09}, in fact, show lumpy structures in SQM ejecta. It is reasonable to assume that further fragmentation and separation of those lumps, as the ejected material approaches its thermodynamic and chemical equilibrium, will ultimately yield a set of strangelets, distributed over a range of mass, that contributes to the primary cosmic rays (PCR)~\cite{bib:mad2005, bib:bis12}. Several other authors have also pointed out the possibility of obtaining strangelets in cosmic rays~\cite{bib:bjorken}. As an aid to the ongoing efforts to detect strangelets in PCR by PAMELA~\cite{bib:casolino}, AMS-02~\cite{bib:kounine} and other experiments, we here determine a plausible mass spectrum of those strangelets at the site of SS merger by invoking a statistical multifragmentation model (SMM) that is often used in the analysis of fragmentation of hot nuclear matter in various contexts~\cite{bib:randrup, bib:gross, bib:bondorf, bib:de07}. This paper is, in fact, a continuation of our earlier attempt~\cite{bib:bis12} to find the basic rate of injection of strangelets of different sizes in the Galaxy. In an exploratory analysis in Ref.~\cite{bib:bis12}, we had ignored quark masses for the sake of simplicity. While this approximation is reasonable for the ``current masses" of the $u$- and the $d$- quarks, ignoring the mass of the $s$-~quarks is questionable. The  value of the current mass ($m_{s}$) of the $s$- quarks had been uncertain for a long time; $m_{s} \sim (100-300)$~MeV had often been considered in the past by several authors~\cite{bib:tar, bib:mad94, bib:barnett95}. After a number of recent high precision estimates~\cite{bib:blossier, bib:particle}, there has now been a consensus on $82~{\rm {MeV}} \lesssim m_{s} \lesssim 100$~MeV, the mostly accepted value being $m_{s} = 95 \pm 5$~MeV~\cite{bib:particle}. 

A finite $m_{s}$ leads to a finite electric charge of the strangelets thus influencing their binding energy through the destabilizing effect of the internal Coulomb repulsion in those strangelets. According to the standard MIT bag model~\cite{bib:farhi, bib:berg1987, bib:gil, bib:mad99}, a large mass of the $s$- quarks makes them ``less relativistic" in comparison with the lighter ($u$- and $d$-) quarks so that the $s$- quarks tend to confine themselves to the interior of the strangelets away from the surface (ie. the boundary) of those strangelets~\cite{bib:gil}. This dynamical property of the massive $s$- quarks causes a depletion of their surface density of states thus contributing a ``quark mode surface tension" to the strangelets; this is often referred to as the ``dynamical surface tension"~\cite{bib:farhi, bib:berg1987, bib:gil, bib:mad99} in the literature. Assuming a chemical potential $\mu_{s} \sim 300$~MeV of the $s$-quarks, the magnitude of this quark mode surface tension is calculated to be $\sim 9~{\rm {MeV~fm}}^{-2}$ at zero temperature. This magnitude is well within the range $\sigma_{s} \sim (5-20)~{\rm {MeV~fm}}^{-2}$ calculated recently in Refs.~\cite{bib:palhares2010} from the linear sigma model (coupled with constituent quarks; LSMq)~\cite{bib:lee} or the Nambu-Jona-Lasinio (NJL) model~\cite{bib:boom}, although the origin of the surface tension is different in those models. Without detailed investigation, one cannot simply rule out the possibility that such quark mode surface tension of the strangelets and their electric charge, that are associated with non-zero values of $m_{s}$~\cite{bib:farhi, bib:mad99}, may make those strangelets unstable (by increasing their energies per baryon above those for ordinary nuclei) so that little or no strangelets may be available in PCR in the solar neighborhood. It is also possible that the fragmentation pattern of strangelets and its variations with various physical parameters, that we found in Ref.~\cite{bib:bis12}, would undergo quantitative or even qualitative changes as the effects of finite $m_{s}$ are taken into account. An examination of these aspects in the fragmentation model presented earlier in Ref.~\cite{bib:bis12} is undertaken in this paper. Fragmentation of color-flavor-locked strange matter (CFL SQM)~\cite{bib:raja01,bib:mad01} will be examined on another occasion in the near future (see also Ref.~\cite{bib:bisposter}). 

The paper is organized along the following line. In Sec.~II, we briefly review the disassembly model originally presented in Ref.~\cite{bib:bis12}. Equations governing the thermodynamic equilibrium of a single strangelet are presented in Sec.~III. In Sec.~IV, we apply the formalism of SMM to find the size distribution of strangelet-fragments. An examination of the stability of the produced fragments, that may be available in PCR in the solar neighborhood, is undertaken in Sec.~V. Discussion of the results and their observational implications are presented in Sec.~VI.    

\section{The multifragmentation model}

\noindent In this model, it is assumed that the initial strange matter, that was tidally released in the merger of two SSs, had rather large (but finite) a volume while the finite-size effects due to its surface and curvature could be ignored. This initial bulk matter (ie.~the SQM) was globally charge-neutral due to the presence of electrons inside the SQM. It is further assumed that the density fluctuations (or fractures), that would eventually lead to fragmentation into strangelets, might have naturally occurred in this warm and excited bulk SQM. The initial average temperature of this bulk matter might have been high enough during its compression in the merger process to allow for such fractures. After it becomes gravitationally unbound from the merged SSs~\cite{bib:bauswein09}, the fractured SQM undergoes quasistatic evolution during which it tries to minimize its free energy by cooling and expanding, while the initial (local) density enhancements develop into more or less well-defined lumps of different baryon numbers (or sizes) still interacting among themselves. The beta-equilibrated and globally charge-neutral lumpy (ie.~fragmenting) matter eventually occupies a freeze-out volume in thermodynamic equilibrium at a certain temperature $T$. It is assumed that, in this volume, both the strong interactions~\cite{bib:bis12} as well as the electromagnetic interactions between the well-developed, (locally) charge-neutral lumps (ie.~the strangelet-fragments in their electron environments) cease to exist. This freeze-out volume is considered to be larger than the original volume of the initially ejected strange matter. The equilibrium temperature $T$ at freeze-out is also considered to be lower than the initial temperature of the tidally released SQM. At the outset, it may appear that, to achieve thermodynamic and chemical equilibrium, the fragmenting system at freeze-out must be at a temperature of about a few tens of MeV which is of the order of the binding energy (per baryon) of the bulk SQM. The fact that such a condition (at thermodynamic equilibrium) is not strictly necessary is perhaps exemplified by the nuclear statistical equilibrium (NSE) established in a pre-bounce, collapsing stellar core in the course of its evolution towards a type II supernova. There, NSE is established at a temperature $\sim 100$~keV~\cite{bib:rosswog} whereas the average binding energy per nucleon of the nuclei is $\sim 8$~MeV. 

In the following, we adapt SMM to find the plausible mass distribution of strangelets in the fragmenting system (at freeze-out) that is often referred to as the ``strangelet-complex" in this paper. This complex is envisaged to be a globally charge-neutral blob of highly inhomogeneous quark matter (accompanied by  electrons), with conserved baryon number $A_{b}$, in its bulk limit; finite-size effects, namely, the surface and the curvature effects of the strangelet-complex are ignored. The blob consists of finite domains that are about to be permanently segregated into numerous sparsely distributed, positively charged, finite-sized strangelets, each of them being embedded in (and possibly penetrated by) a charge-neutralizing cloud of degenerate electrons having a volume that is much larger than the volume of the strangelet. At freeze-out, the average distance between the strangelets is large such that the residual strong interactions between those strangelets can be considered to be insignificant. The globally charge-neutral strangelet-complex, comprising of the positively charged strangelets immersed in an  electron gas, obtained by considering all the electron clouds together, is in thermodynamic and chemical equilibrium (including beta-equilibrium) at freeze-out. Thermodynamic equilibrium ensures that the temperature ($T$) is constant throughout the strangelet-complex. The number density of electrons in this complex has spatial variation ($n_{e}^{i}(r)$), with length scale comparable to (or even shorter than) the Debye screening length $\lambda_{D} = (\frac{\pi}{8\alpha})^{1/2}\frac{1}{\mu_{q}} \approx \frac{7.33}{\mu_{q}}\sim 5$~fm~\cite{bib:hei93, bib:jensen, bib:alford, bib:alford08, bib:alford2012} inside each individual fragment; $r$ being the radial distance from the centre of an (assumed spherical) strangelet-fragment of the $i^{\rm {th}}$ species characterized by its baryon number $A^{i}$. Here, $\mu_{q}~(\gg T)$ is the quark number chemical potential~\cite{bib:alford2002} of the strangelet-complex in thermodynamic equilibrium at freeze-out and $\alpha$ ($= \frac{e^{2}}{\hbar c}$) is the fine structure constant, $e$ being the magnitude of the electronic charge, $c$ is the speed of light and $\hbar = h/2 \pi$ with $h$ being the Planck's constant. The chemical potential $\mu_{q}$ is equal to one third of the baryon number chemical potential~\cite{bib:alford2002} of the strangelet-complex; thermodynamic equilibrium at constant baryon number demands that $\mu_{q}$ is constant over the volume of that complex. Except for its short-scale spatial variations as stated above, the distribution of electrons (as viewed over length scales that are orders of magnitude longer than $\lambda_{D}$) in the strangelet-complex should otherwise be uniform, necessitated by the conditions of global equilibrium, with constant number density $n_{e}$. As a consequence, the chemical potential of the electrons $\mu_{e} \approx (3\pi^{2}n_{e})^{1/3}$ is also a global constant, satisfying $\mu_{e} \lesssim \frac{m_{s}^{2}}{4 \mu_{q}}$, over the volume of the strangelet-complex. Here, the upper limit for $\mu_{e}$ corresponds to the electron chemical potential in a globally charge-neutral, cold ($T=0$) and uniform bulk quark matter at zero pressure~\cite{bib:alford, bib:alford08, bib:alford2012, bib:jaikumar2006, bib:alford2002}. This limit is derived by assuming $\mu^{2}_{q} \gg m^{2}_{s}$ that is justified in the case $m_{s} \approx 95$~MeV and $\mu_{q} \sim 300$~MeV. In our calculations, that aim only for order of magnitude accuracies, we would consider $\mu_{e} \approx \frac{m_{s}^{2}}{4 \mu_{q}}$ for the sake of simplicity. 

The condition of chemical equilibrium of the strangelet-complex, that is maintained by the weak interaction processes $d \leftrightarrow u + e + \bar{\nu_{e}}$, $s \leftrightarrow u + e + \bar{\nu_{e}}$ and $s + u \leftrightarrow u + d$, demands that the chemical potential $\mu_{f}$ of the quarks of the $f^{\rm {th}}$ flavor~($f = u, d, s$) in the strangelet-complex should satisfy the relations~\cite{bib:alford2002}:

\begin{equation}
\begin{rcases}
\mu_{u} &= \mu_{q} - \frac{2}{3}\mu_{e} \\
\mu_{d} &= \mu_{q} + \frac{1}{3}\mu_{e} \\
\mu_{s} &= \mu_{q} + \frac{1}{3}\mu_{e}
\end{rcases}
\text{}
\end{equation}

\noindent  thus implying that, except for its local variation ($\mu_{f}^{i}(r)$) with length scale~$\lesssim \lambda_{D}$ inside each individual strangelet-fragment, the value of the chemical potential $\mu_{f}$ of an arbitrary quark-flavor ($f$) should otherwise be the same throughout the volume of the strangelet-complex.

In Eqs.~(1), we ignored the chemical potential of the neutrinos as they are likely to contribute very little to the energy density and pressure of the strangelet-complex in equilibrium at freeze-out~\cite{bib:farhi}. We also note that, over length scale~$\lesssim \lambda_{D}$ inside a strangelet of the $i^{\rm {th}}$ species, the equilibrium conditions in Eq.~(1) transform into a relation between the local (ie. position dependent) chemical potentials, ie.

\begin{equation}
\mu_{f}^{i}(r) + q_{f}\mu_{e}^{i}(r) = \mu_{q};
\end{equation}

\noindent $q_{f}$ being the charge of a quark of the $f^{\rm {th}}$ flavor, ie. ($q_{u}$, $q_{d}$, $q_{s}$) $= (\frac{2}{3}, \frac{-1}{3}, \frac{-1}{3})$ in the unit of $e$. Eq.~(2) couples the equilibrium distribution of quarks inside strangelets to the charge-neutralizing electron clouds surrounding those strangelets through non-zero values of the local electrostatic potentials $\mu_{e}^{i}(r)/e$~\cite{bib:alford, bib:alford08, bib:alford2012} at different positions inside strangelets. This coupling gives rise to the phenomenon of Debye screening  ~\cite{bib:hei93, bib:jensen, bib:alford, bib:alford08, bib:alford2012} inside a strangelet in thermodynamic, electrostatic and chemical equilibrium with its electron environment. In Sec.~III of this paper, we will use the formulae, obtained by earlier authors~\cite{bib:hei93, bib:jensen, bib:alford}, that express the values of two integrated (over the radial coordinate $r$) properties, namely, the total electric charge and the Coulomb energy of an individual fragment in the strangelet-complex. The derivations of those formulae make use of the concept of Debye-screening inside an individual fragment. It is, however, important to note that, in this paper, our purpose is to describe the equilibrium size-distribution of numerous new-born strangelets (each embedded in an electron cloud) located randomly within the freeze-out volume of the strangelet-complex. The chemical potentials that we use for this purpose are the ones (ie. $\mu_{e}$, $\mu_{f}$ and $\mu_{q}$) characterizing the global equilibrium configuration of the strangelet-complex (at freeze-out) instead of the position dependent local chemical potentials ($\mu^{i}_{f}(r)$ and $\mu^{i}_{e}(r)$) of the quarks and electrons pertaining to each individual fragment. The latter potentials enter in our analysis only indirectly, ie. through the integrated properties of each individual strangelet mentioned above. In principle, these global and local chemical potentials should be connected through appropriate boundary conditions on the surface of each individual strangelet. Examples of these boundary conditions were provided in Refs.~\cite{bib:alford, bib:alford08, bib:alford2012} in which rigorous calculations (in the form of the solutions of separate Poisson's equations inside and outside the strangelet that match on its surface) were undertaken to numerically determine the equilibrium (radial) distributions of charge density and other physical properties inside and outside an individual strangelet embedded in an inhomogeneous electron cloud. Those calculations also provided us with a physical insight into the phenomenon of Debye screening. Determination of such detailed internal structure of each individual strangelet in equilibrium with its electron environment is, however, not attempted in this paper.

  In the above, we have considered the strangelet-complex to be in thermodynamic equilibrium at a certain temperature ($T$) at freeze-out.  In real SS merger events, there may, however, be a distribution of those temperatures in the regions in which strangelets are formed. In the absence of observations or the results from the relevant numerical simulations, it is difficult to definitely predict a range of temperatures of those regions. Here, we arbitrarily consider the values of $T$ within a range ($0.001-1.0$)~MeV at freeze-out. The lower limit ($\sim 1$~keV) corresponds to the possible lower bound of the temperatures usually associated with the accretion disks of the low mass X-ray binary systems~\cite{bib:rosswog}. The upper limit ($\sim 1$~MeV) corresponds to the magnitude of temperatures attained by the materials ejected from the tip of the tidal arms formed sometimes in the simulations~\cite{bib:oechslin} of merger between two neutron stars (NSs). 

At freeze-out, the multiplicity $\omega^{i}$ of the strangelets of species `$i$' is written as~\cite{bib:bondorf}

\begin{equation} 
\omega^{i} = \frac{{\cal V}}{({\cal L}^{i})^{3}}e^{(\mu^{i} - F^{i})/T}.
\end{equation}

\noindent Here, $\cal V$ is the available volume, ie. the freeze-out volume minus the volume of the produced fragments. In Eq.~(3), $\mu^{i} (= \sum_{f} \mu_{f} N^{i}_{f} + \mu_{e}N^{i}_{e})$ is the chemical potential of a strangelet of the $i^{\rm th}$ species having a volume ${\sf{V}}^{i}$ with $N^{i}_{(f,e)}$ = $\Big(-\frac{\partial \Omega^{i}_{(f,e)}}{\partial \mu_{(f,e)}}\Big)_{{\sf{V}}^{i}, T}$ being either the number of quarks of the $f^{\rm {th}}$ flavor ($f = u,d,s$) or the number of electrons constituting that strangelet; their corresponding thermodynamic potentials are $\Omega^{i}_{(f,e)}$. The thermal de-Broglie wavelength of a strangelet of the $i^{\rm {th}}$ species is defined as ${\cal L}^{i} = h/\sqrt{2\pi m^{i}T}$ where $m^{i}$ is the mass of the strangelet. For an approximate value of $m^{i}$, we consider the mass-formulae derived in Refs.~\cite{bib:mad95, bib:mad99} by using a bulk approximation to the baryon number chemical potential of the strangelet-fragment at $T = 0$. The mass of a strangelet with $m_{s} = 95$~MeV is obtained by means of an interpolation between the masses derived in Ref.~\cite{bib:mad95} for different values of $m_{s}$. The masses of strangelets corresponding to various bag values are obtained by using the scaling law derived in Ref.~\cite{bib:mad99}. Here, $F^{i} ( = \Omega^{i} + \mu^{i} + E^{i}_{\rm C})$ stands for the Helmholtz free energy of the $i^{\rm {th}}$ species while $\Omega^{i}$ is its thermodynamic potential and $E^{i}_{\rm C}$ is its Coulomb energy. $F^{i}$ may be rewritten as $F^{i} = {\Omega^{i}}_{\rm {tot}} + \mu^{i}$, where, $\Omega^{i}_{\rm {tot}} = \Omega^{i} + E^{i}_{\rm C}$. Thus, Eq.~(3) can be reframed as

\begin{equation}
\omega^{i} = \frac{{\cal V}}{({\cal L}^{i})^{3}}e^{-\Omega^{i}_{\rm {tot}}/T}.
\end{equation}

\noindent We will use Eq.~(4) to determine the multiplicities of various fragments in the strangelet-complex after specifying the thermodynamic quantities representing the overall behaviour of an individual strangelet in that complex in equilibrium at freeze-out. We also add that, throughout this work, we choose natural units such that $\hbar = c = k_{\rm B} =1~{\rm {and}}~\alpha = \frac{1}{137}$, where $k_{\rm B}$ is the Boltzmann constant. Surface tension and curvature coefficient of a strangelet-fragment (see the next section) are expressed in the units of ${\rm {MeV~fm}}^{-2}$ and ${\rm {MeV~fm}}^{-1}$, respectively. Any number density appearing in this paper is either in the unit of $({\rm {fm}})^{-3}$ or in the unit of $({\rm {MeV}})^{3}$, to be stated explicitly.

\section{Thermodynamics of a strangelet}

\noindent To calculate the thermodynamic potential of the strangelet of a particular species, we shall use the multiple reflection expansion method~\cite{bib:balian} with smoothed density of states as applied to the standard MIT bag model~\cite{bib:farhi,bib:mad95,bib:mad99}. This method is similar to the liquid drop model used in the theories of nuclear structure~\cite{bib:mad99}. It is known that the model can satisfactorily fit with the average properties of the strangelets that are obtained from the mode-filling calculations of the shell model~\cite{bib:mad94, bib:mustafa1996}. The strength of the QCD coupling between quarks is taken to be zero here; it was argued that the effect of such coupling may be absorbed by a rescaling of the bag constant~\cite{bib:farhi, bib:mad99}. The strangelets are assumed to be spherical in shape for the sake of simplicity; thermodynamic properties of a deformed strangelet at finite temperature was discussed in Ref.~\cite{bib:mustafa1997}. Radius of a spherical strangelet of the $i^{\rm {th}}$ species is $R^{i} = r^{i}_{\rm o}(A^{i})^{1/3}$, $r^{i}_{\rm o}$ being its radius parameter. Volume, surface and curvature of a strangelet are denoted as ${\sf{V}}^{i} = \frac{4}{3}\pi (R^{i})^{3}$, ${\sf{S}}^{i} = 4\pi (R^{i})^{2}$ and ${\sf{C}}^{i} = 8\pi R^{i}$, respectively. Thermodynamic potential of a strangelet of the $i^{\rm {th}}$ species is written as

\begin{equation}
\Omega^{i} = \sum_{f}\Omega^{i}_{f} + \Omega^{i}_{e} + \Omega^{i}_{\rm {gluon}} + B{\sf V}^{i} = \Omega_{\sf V}^{\rm o} {\sf V}^{i} + \Omega_{\sf S}^{\rm o} {\sf S}^{i} + \Omega_{\sf C}^{\rm o} {\sf C}^{i} + B{\sf V}^{i},
\end{equation}

\noindent where, the contribution $\Omega^{i}_{\rm {gluon}}$ is obtained from Ref.~\cite{bib:mad99}. In Eq.~(5),

\begin{widetext}
\begin{subequations}
\begin{eqnarray}
\Omega_{\sf V}^{\rm o} = -\frac{37}{90}\pi^{2}T^{4} - \Big(\frac{\mu_{u}^{2}+\mu_{d}^{2}}{2}\Big)T^{2} - \Big(\frac{\mu_{u}^{4}+\mu_{d}^{4}}{4 \pi^{2}}\Big) - \frac{\mu^{4}_{s}}{4\pi^{2}}\Bigg[\Big(1 - \frac{5}{2}\lambda_{s}^{2}\Big)\sqrt{1 - \lambda_{s}^{2}} + \frac{3}{2}\lambda^{4}_{s}\ln\Bigg(\frac{1 + \sqrt {1 - \lambda_{s}^{2}}}{\lambda_{s}}\Bigg)\nonumber \\
+ 2\pi^{2}\Big(\frac{T}{\mu_{s}}\Big)^{2}\sqrt {1 - \lambda_{s}^{2}} + \frac{7\pi^{4}}{15}\Big(\frac{T}{\mu_{s}}\Big)^{4}\frac{(1 - \frac{3}{2}\lambda_{s}^{2})}{(1 - \lambda_{s}^{2})^{3/2}}\Bigg] - \frac{\mu_{e}^{4}}{12 \pi^{2}},\nonumber \\
~~~~~~~~~~~~~~~~~~~~~~~~~~~~~~~~~~~~~~~~~~~~~~~~~~~~~~~~~~~~~~~~~~~~~~~~~~~~~~
\end{eqnarray}

\begin{eqnarray}
\Omega_{\sf S}^{\rm o} = \frac{3}{4 \pi}\mu_{s}^{3}\Bigg[\frac{(1-\lambda^{2}_{s})}{6} - \frac{\lambda_{s}^{2}}{3}(1-\lambda_{s}) - \frac{1}{3\pi}\Bigg\{\tan^{-1}{\Bigg(\frac{\sqrt{1-\lambda^{2}_{s}}}{\lambda_{s}}\Bigg)} + \lambda^{3}_{s} \ln\Bigg(\frac{1 + \sqrt{1-\lambda^{2}_{s}}}{\lambda_{s}}\Bigg) -2\lambda_{s}\sqrt{1-\lambda^{2}_{s}}\Bigg\} \nonumber\\
+ \frac{\pi}{3}\Big(\frac{T}{\mu_{s}}\Big)^{2}\Bigg\{\frac{\pi}{2} - \tan^{-1}{\Bigg(\frac{\sqrt{1-\lambda^{2}_{s}}}{\lambda_{s}}\Bigg)}\Bigg\} + \frac{7 \pi^{3}}{180}\Big(\frac{T}{\mu_{s}}\Big)^{4}\frac{\lambda^{3}_{s}}{(1-\lambda^{2}_{s})^{3/2}}\Bigg]\nonumber \\
~~~~~~~~~~~~~~~~~~~~~~~~~~~~~~~~~~~~~~~~~~~~~~~~~~~~~~~~~~~~~~~~~~~~~~~~~~~~~~~~~~~~~~~
\end{eqnarray}

\noindent and

\begin{eqnarray}
\Omega_{\sf C}^{\rm o} = \frac{19}{36}T^{2} + \Big(\frac{\mu_{u}^{2} + \mu_{d}^{2}}{8 \pi^{2}}\Big)
+ \frac{\mu^{2}_{s}}{8 \pi^{2}}\Bigg[\frac{1}{\lambda_{s}}\Bigg\{\frac{\pi}{2} - \tan^{-1}{\Bigg(\frac{\sqrt{1-\lambda^{2}_{s}}}{\lambda_{s}}\Bigg)} \Bigg\} + \Big(\frac{\pi^{2}}{\lambda_{s}}\Big)\Big(\frac{T}{\mu_{s}}\Big)^{2}\Bigg\{\frac{\pi}{2} - \tan^{-1}{\Big(\frac{\sqrt{1-\lambda^{2}_{s}}}{\lambda_{s}}\Big)}\Bigg\} \nonumber\\
+ \lambda^{2}_{s}\Bigg\{\pi + \ln{\Bigg(\frac{1+\sqrt{1-\lambda^{2}_{s}}}{\lambda_{s}}\Bigg)}\Bigg\} -\frac{3\pi}{2}\lambda_{s} - \Big(\frac{2\pi^{2}}{3}\Big)\Big(\frac{T}{\mu_{s}}\Big)^{2}\frac{1}{\sqrt{1-\lambda^{2}_{s}}}  
 - \frac{7\pi^{4}}{60}\Big(\frac{T}{\mu_{s}}\Big)^{4}\frac{\lambda^{2}_{s}(1+\lambda^{2}_{s})}{(1-\lambda_{s}^{2})^{5/2}}\Bigg] \nonumber\\
~~~~~~~~~~~~~~~~~~~~~~~~~~~~~~~~~~~~~~~~~~~~~~~~~~~~~~~~~~~~~~ 
\end{eqnarray}
\end{subequations}
\end{widetext}

\noindent with $\Omega_{\sf V}^{\rm o}$, $\Omega_{\sf S}^{\rm o}$ and $\Omega_{\sf C}^{\rm o}$ being the thermodynamic potential densities associated with the volume, surface and curvature of the strangelets. The quantity $\Omega_{\sf S}^{\rm o}$ is readily recognizable as the quark mode surface tension while the quantity $\Omega_{\sf C}^{\rm o}$ is defined as the curvature coefficient of the strangelet. In Eqs.~(6), $\lambda_{s} = \frac{m_{s}}{\mu_{s}}$ and $B$ is the bag pressure. In case of the assumption $m_{s} \rightarrow 0$, we obtain $\lambda_{s} \rightarrow 0$ and $\mu_{u} = \mu_{d} = \mu_{s} = \mu_{q}$. In this limit of vanishingly small quark-masses, charge-neutrality of the strangelet-complex no longer requires the presence of electrons in that complex so that $\mu_{e} \rightarrow 0$ as $m_{s} \rightarrow 0$. Eqs.~(5) and (6) reduce to the thermodynamic potential of an isolated (ie. not embedded in an electron cloud) strangelet in the limit of massless quarks as given in Refs.~\cite{bib:mad99,bib:bis12} in this situation. For $T =0$, the thermodynamic potential of cold strangelets with massive $s$-~quarks~\cite{bib:mad99} is restored from Eqs.~(5) and (6) except for an additional term involving $\mu_{e}^{4}$ in Eq.~(6a). This term was ignored in Ref.~\cite{bib:mad99} while discussing the thermodynamics of a small, isolated strangelet not embedded in a charge-neutralizing background of electrons so that $\mu_{e} = 0$ for that strangelet. In this paper, we take the presence of those electrons into account.

It is apparent from Eqs.~(5) and (6b) that the quark mode surface energy, proportional to $({R^{i}})^{2}$, vanishes in the limit $m_{s} \rightarrow 0$ in the   traditional MIT bag model of strangelets~\cite{bib:farhi, bib:berg1987}. Sophisticated models, such as the LSMq and the NJL models (see Sec.~I), of SQM, however, predict appreciable surface tension (at the vacuum-quark matter phase boundary) even in this case of massless quarks constituting the SQM~\cite{bib:palhares2010}. This absence of surface tension in the zero quark-mass, traditional MIT bag model seems to be a direct consequence of its ignoring the effects of dynamical (or explicit) chiral symmetry breakdown in the QCD vacuum. According to the Nambu-Goldstone theory of chiral symmetry, such a breakdown would lead to a qualitative re-arrangement of the QCD vacuum by enabling it to host strong condensates of quark-antiquark pairs~\cite{bib:pokrovskii1989, bib:weise, bib:hartmann1999}. The discontinuities of those condensates on the bag surface would lead to a surface tension determined by the sum of the quark condensates~\cite{bib:pokrovskii1989}. Notwithstanding this serious shortcoming (ie. its failure to take the consequences of dynamical (or explicit) breaking of QCD chiral symmetry into consideration) of the traditional MIT bag model, we still employ this model in the present paper for its mathematical and conceptual simplicity in modelling the consequences of selected features (in QCD), namely the short distance ($<0.1~{\rm {fm}}$) asymptotic freedom as well as a perfect spatial and color confinement of the quarks at long ($>1~{\rm {fm}}$) distance scales by the bag pressure ($B$), which is assumed (in this model) to include all the non-perturbative effects from the QCD vacuum on the quarks inside the bag~\cite{bib:hartmann1999}. As we demonstrated earlier in Ref.~\cite{bib:bis12}, the above simplicity of the standard MIT bag model proves to be convenient for a preliminary investigation of the multifragmentation of SQM. The mathematically harder tasks of describing multifragmentation in more sophisticated theoretical models of SQM, such as the LSMq and the NJL models or the chirally invariant bag models~(eg.~\cite{bib:pokrovskii1989, bib:bhaduri}), will be attempted in near future.

While the traditional MIT bag model (with massless quarks) is marked by the absence of a surface tension, it nevertheless provides for a positive curvature coefficient (Eq.~(6c)) $\sigma_{c} = \Omega_{\rm C}^{\rm o} = (3/8\pi^{2}){\mu_{q}}^{2} \sim 17~{\rm {MeV~fm}}^{-1}$ (at zero temperature and for $\mu_{q} \sim 300~{\rm {MeV}}$)~\cite{bib:mad1993, bib:mad99} of the strangelets. This curvature coefficient arises from finite-size corrections to the quark density of states that are required to match the quark wave functions to the bag boundary conditions~\cite{bib:mad1993}. In analogy with the quark mode surface tension (in Sec.~I), we may refer to this (curvature) coefficient as the ``quark mode curvature coefficient" of the strangelets. In spite of the important distinction between the energies associated with them (which scale differently with the baryon number $A^{i}$; see Eq.~(5)), the surface tension and the curvature coefficient play somewhat similar roles in the multifragmentation of SQM. Both these quantities require additional energy to produce copious smaller fragments at the expense of a few large fragments as we will find in Sec.~IV. Moreover, both the surface tension and the curvature coefficient tend to destabilize finite-sized fragments by increasing their energies per baryon; see Sec.~V.  In view of the above, we may like to investigate into the relative magnitude of the quark mode curvature energy (in the limit of massless quarks in the MIT bag model) of the strangelets {\textit{vis-a-vis}} their surface energy determined in Refs.~\cite{bib:palhares2010} from the LSMq and the NJL models as mentioned in Sec.~I. For a strangelet having $A^{i} \sim 10$, the value of the curvature energy (per baryon) turns out to be $\sigma_{c}{\sf{C}}^{i}/A^{i}  \sim 100~{\rm MeV}$, whereas, the value of the surface energy (per baryon) determined in Refs.~\cite{bib:palhares2010} lies in the range~$\sigma_{s}{\sf{S}}^{i}/A^{i}\sim (30-120)~{\rm {MeV}}$. For $A^{i} \sim 100$, the values of these energies are $\sim 20~{\rm {MeV}}$ and $\sim (10-50)~{\rm {MeV}}$, respectively. Similarly, these quantities take on values $\sim 4~{\rm {MeV}}$ and $\sim (6-25)~{\rm {MeV}}$ for $A^{i}\sim 10^{3}$. In the above, we have considered $m_{s} = 0$, $r^{i}_{\rm o} \sim 1~{\rm {fm}}$~\cite{bib:hei93, bib:he}, $T = 0$ and $\mu_{q} \approx 300~{\rm {MeV}}$ to calculate the curvature energies of strangelets by using the MIT bag model. Above comparisons reveal that, for an approximate range of most of the fragment-sizes ($10 \lesssim A^{i} \lesssim 10^{3}$) obtained (in Sec.~IV) in this paper, the magnitude of curvature energy of strangelets (with massless quarks in the MIT bag model) is more-or-less compatible at least with the lower bound of the surface energy determined in Refs.~\cite{bib:palhares2010}. Numerical examples presented above seem to suggest that, notwithstanding the absence of a surface energy for massless quarks in the MIT bag model, it is perhaps not unreasonable to use this model (having a positive curvature coefficient) for the sake of a rudimentary analysis of multifragmentation of SQM. Such an analysis, undertaken originally in Ref.~\cite{bib:bis12}, may provide us with some basic idea that would possibly be useful in the computations of multifragmentation in the sophisticated models of SQM. We add here that the present paper, that employs MIT bag model with $m_{s} \ne 0$, is relatively less vulnerable to the above drawback as, along with a positive curvature coefficient, the model also predicts a positive quark mode surface tension whose magnitude is found (in Sec.~I) to be compatible with the ones in Refs.~\cite{bib:palhares2010}. In our calculations, we do not get any negative curvature coefficient as obtained in Refs.~\cite{bib:christ1997} in the (theorized) cases of the quark-hadron phase boundary and the kaon condensate-normal nuclear matter phase boundary in neutron stars. We may however note that, according to the authors of Ref.~\cite{bib:haen2007}, such a negative curvature coefficient may, as well, appear due to certain oversimplified assumptions used in the calculations.

In Eqs.~(5) and (6), we have taken account of the fact that $\mu_{e} \ll \mu_{f}; f = (u, d, s)$. We, therefore, considered only the leading order contribution from the electron chemical potential in $\Omega_{\sf V}^{\rm o}$ in Eq.~(6a). We further assume that, even though the electron chemical potential is non-zero inside strangelets, the sizes of those strangelets are too small to have electrons localized inside them, so that, the contribution of those electrons to the thermodynamic potentials associated with the surface and the curvature of a strangelet (Eqs.~(6b) and (6c)) need not be taken into account. The electric charge of a strangelet is given by that of the quark matter alone in the case of such small strangelets~\cite{bib:alford}. Moreover, as the $s$- quarks are massive, the number of those quarks is less than the numbers of quarks of other (viz.~$u$-~and $d$-~) flavors inside the strangelet. A strangelet of the $i^{\rm {th}}$ species possesses a positive charge number $Z^{i}$ in this situation~\cite{bib:farhi, bib:alford, bib:hei93}. It was pointed out in Refs.~\cite{bib:farhi, bib:hei93, bib:berg1987} that the approximation of no localized electrons inside strangelets may be justified when $A^{i} \lesssim 10^{5}$ so that the radius of a strangelet satisfies the condition $R^{i} \lesssim 46~{\rm {fm}} < a_{\rm{B}}/Z^{i} \sim 253~{\rm {fm}} < 2 \pi/m_{\rm e} \sim 2.4\times 10^{3}~{\rm {fm}}$. Here, $a_{\rm B} = 1/(\alpha m_{\rm e})$ is the Bohr radius; $m_{\rm e}$ is the mass of the electron and $2 \pi/m_{\rm e}$ is the electron Compton wavelength in the units considered in this paper. In the above, $\mu_{u} \sim \mu_{d} \sim \mu_{s} \sim \mu_{q} \sim 300$~MeV~\cite{bib:farhi, bib:hei93} and $Z^{i} \leq 214$~(see Eq.~(7) below). We have checked that the above condition for the absence of localized electrons inside strangelets is satisfied by all the fragments in the strangelet-complex that we finally obtain in Sec.~IV. The charge-neutralizing electron cloud surrounding such a strangelet has been treated here within the framework of the Wigner-Seitz (WS) approximation as described in a later paragraph of this section.

In this paper, we take the effect of Debye screening~\cite{bib:hei93, bib:jensen, bib:alford, bib:alford08, bib:alford2012} on the charge distribution inside relatively larger ($a_{\rm B}/Z^{i}> R^{i} > \lambda_{\rm D}$)~strangelets into account. Equilibrium of the quarks in the electrostatic field inside such a strangelet bar its core from having a positive electric charge density, ie. the deep interior of that strangelet is charge neutral; see the discussion in Sec.~II above. The positive charge density inside that strangelet is confined within a layer of thickness $\sim \lambda_{D}$ from its surface. In this situation, the total charge and the Coulomb energy of the strangelet are obtained by integrating over the radial coordinate ($r$) measured from its centre. The expressions of these integrated quantities are given as~\cite{bib:hei93, bib:jensen, bib:alford}:

\begin{equation}
Z^{i} \approx \frac{m_{s}^{2}}{4 \alpha \mu_{q}} R^{i} \Bigg[1 - \frac{\tanh (R^{i}/\lambda_{D})}{(R^{i}/\lambda_{D})}\Bigg] 
\end{equation} 

\noindent and 
\begin{eqnarray}
E^{i}_{\rm C} \approx \frac{m_{s}^{4}}{32 \alpha \mu_{q}^{2}} R^{i} \Bigg[1 - \frac{3}{2}\frac{\tanh(R^{i}/\lambda_{D})}{(R^{i}/\lambda_{D})} \nonumber \\
+ \frac{1}{2}\Bigg\{\cosh (R^{i}/\lambda_{D})\Bigg\}^{-2}\Bigg].
\end{eqnarray} 

\noindent In reality, after the application of SMM to the initial bulk matter, we would have rather large an array of strangelets of various sizes. Many of those strangelets may not be large enough to satisfy the condition of charge-screening. Whatever may be the case, Eqs.~(7) and (8) are generalized enough to equally account for the large ($R^{i} > \lambda_{D}$) as well as the small ($R^{i} \lesssim \lambda_{D}$) strangelets. In the following, we, therefore, adapt those two equations to proceed with the calculations. 
  
The entropy of a strangelet of the $i^{\rm {th}}$ species is ${\cal S}^{i} = - \Big(\frac{\partial \Omega^{i}}{\partial T}\Big)_{{\sf V}^{i}, \mu^{i}}$. Thus, the total energy of a strangelet may be written as 
\begin{eqnarray}
E^{i} = &&T{\cal S}^{i} + \mu^{i} + \Omega^{i}_{\rm {tot}} \nonumber \\
      = &&T{\cal S}^{i} + \mu^{i} + \Omega^{i} + E^{i}_{\rm C}.
\end{eqnarray}

For thermodynamic equilibrium, the strangelet-fragments, in addition to being in electrostatic and chemical equilibrium (including beta-equilibrium) are also in mechanical equilibrium, ie. $P^{i}_{\rm {ext}} = - \Big(\frac{\partial \Omega^{i}_{\rm {tot}}}{\partial \sf {V}^{i}}\Big)_{T, \mu^{i}}$; $P^{i}_{\rm {ext}}$ being the external pressure (as distinct from the bag pressure) on a strangelet of the $i^{\rm {th}}$ species. This pressure is assumed to be exerted by $Z^{i}$ charge-neutralizing relativistic electrons residing outside the $i^{\rm {th}}$ fragment but within the WS cell surrounding that particular fragment. Following Ref.~\cite{bib:salpeter61}, a strangelet-fragment of the $i^{\rm {th}}$ species is approximated to be a pointlike (ie. ${\sf V}^{i} \ll V^{i}_{\rm {cell}}$; $V^{i}_{\rm {cell}}$ being the volume of the $i^{\rm {th}}$ WS cell) positive charge ($Z^{i}$) surrounded by the spherical WS cell containing electrons of uniform number density (see the discussion preceding Eq.~(1) in Sec.~II) $\frac{Z^{i}}{V^{i}_{\rm {cell}}} = \frac{N^{\rm {total}}_{e}}{\cal V} = n_{e} \approx \frac{m_{s}^{6}}{192 \pi^{2} \mu_{q}^{3}}$;  $N^{\rm {total}}_{e}$ being the total number of electrons in the strangelet-complex. We choose $V^{i}_{\rm {cell}} = \frac{Z^{i}}{{\sum_{i}Z^{i}\omega^{i}}}{\cal V}$ so that it satisfies the condition $\sum_{i}{\omega^{i}V^{i}_{\rm {cell}}} = \cal V$. The expression for the pressure of relativistic electrons on the strangelet may then be written as~\cite{bib:salpeter61}

\begin{eqnarray}
&&P^{i}_{\rm {ext}} \approx ({3 \pi^{2}})^{1/3} \Big[\frac{({{n_{\rm e}}})^{4/3}}{4}\Big] + \Big(\frac{\pi^{2}}{2}\Big) \Big[\frac{{T^{2}}}{({3 {\pi^{2}}})^{1/3}}\Big] ({n_{\rm e}})^{2/3}\nonumber \\
&&- \Big(\frac{3}{10}\Big)\Big(\frac{4\pi}{3}\Big)^{1/3}\alpha(Z^{i})^{2/3}({n_{e}})^{4/3} \nonumber \\
&&- \Big(\frac{1}{6}\Big)\Big(\frac{324}{175}\Big)\Big(\frac{4}{9\pi}\Big)^{2/3} (3 \pi^{2})^{1/3} (Z^{i})^{4/3} \alpha^{2} (n_{e})^{4/3}\nonumber \\
&&+ \Big(\frac{1}{8\pi}\Big)\alpha (3\pi^{2})^{1/3} ({n_{e}})^{4/3}
-\frac{0.062}{6}\alpha^{2} m_{e} n_{e}.
\end{eqnarray}
    
\noindent In Eq.~(10), the first two terms on the right hand side represent the pressure of a degenerate Fermi gas of noninteracting, relativistic electrons at  temperature $T$. The third term stands for the Coulomb interactions between the pointlike strangelet and the uniformly distributed electrons as well as the electron-electron interactions. The fourth term in Eq.~(10) represents the Thomas-Fermi correction that results from first order deviation of the electron distribution from uniformity. This deviation is obtained by expanding the relativistic electron kinetic energy about its value given by the uniform approximation and then assuming that the ratio of the Coulomb potential energy of the electron to the electron Fermi energy to be of the same order as the deviation in the electron distribution. The fifth term arises due to the interactions between the relativistic electrons \textit{via.}~transverse electromagnetic field while the sixth term represents the influence of the electric field of ions on the above interactions between electrons. After comparing with a more rigorous treatment of the WS cell~\cite{bib:rotondo2011} containing electrons around a finite-sized nucleus by the use of a relativistic generalization of the Feynman-Metropolis-Teller algorithm that also takes the penetration of electrons inside the nucleus into account, we found that the approximation in Ref.~\cite{bib:salpeter61} is satisfactory in the case of strangelet species of relatively large multiplicities for which significant fluxes may be expected in the neighborhood of the Sun; such multiplicities of strangelets are determined in Sec.~IV and their fluxes in the vicinity of the solar system are estimated in Sec.~VI of this paper.

Here, it is important to note that an analytical expression for the electron pressure, that is similar to the one in Eq.~(10), was obtained earlier in Ref.~\cite{bib:alford2012} by using a ``low pressure approximation" in which the electrons may be assumed to have an uniform number density inside the WS cell pertaining to a strangelet. In Ref.~\cite{bib:alford2012}, the said approximation was found to be justified in the case $a_{\rm B}/Z^{i}> R^{i}_{\rm {cell}}\gg R^{i}$ with $R^{i}_{\rm {cell}} = (\frac{3}{4\pi}V^{i}_{\rm {cell}})^{1/3}$ being the radius of the spherical WS cell. We have verified that the above criterion is satisfied by all the WS cells associated with the charged fragments (in the strangelet-complex) obtained numerically in Sec.~IV by using the relations $\mu_{e} \approx \frac{m^{2}_{s}}{4\mu_{q}}$ and $n_{e} \approx \frac{\mu^{3}_{e}}{3 \pi^{2}}$ mentioned in Sec.~II. The authors of Ref.~\cite{bib:alford2012} determined the pressure of the charge-neutralizing electrons on a strangelet by considering the contributions from the degeneracy pressure (at zero-temperature) of electrons along with that from the Coulomb interactions between the strangelet and the electrons as well as a contribution from the finite electron-mass ($m_{e}$) within the framework of the low pressure approximation described above. We have checked that the discrepancy between the numerical values of multiplicities of strangelets, obtained (in Sec.~IV) by using Eq.~(10) and then by using the expression in Ref.~\cite{bib:alford2012}, is not more than about $.01\%$ for the ranges of values of temperature, bag parameter and fragment-size considered in this paper.

With the definitions given in Eqs.~(5), (6), (7), (8) and (10), the condition for mechanical equilibrium mentioned above ultimately yields an expression for the total thermodynamic potential $\Omega^{i}_{\rm {tot}}$ (defined in the discussion preceding Eq.~(4) in Sec.~II) of the strangelet at thermodynamic equilibrium at freeze-out. This expression is

\begin{widetext}
\begin{equation}
\Omega^{i}_{\rm {tot}} = (-\Omega^{\rm o}_{\sf V} - B){\sf V}^{i}\Big(\frac{\Omega^{\rm o}_{\sf S}{\sf S}^{i} + 2\Omega^{\rm o}_{\sf C}{\sf C}^{i} + 3E^{i}_{\rm C}  -\Delta E^{i}_{C} - 3P^{i}_{\rm {ext}}{\sf V}^{i}}{2\Omega^{\rm o}_{\sf S}{\sf S}^{i} + \Omega^{\rm o}_{\sf C}{\sf C}^{i} + \Delta E^{i}_{C} + 3 P^{i}_{\rm {ext}}{\sf V}^{i}}\Big),
\end{equation}

\noindent where,

\begin{equation}
\Delta E^{i}_{C} \approx \frac{m_{s}^{4}}{32 \alpha \mu_{q}^{2}}R^{i}\Bigg[1 - \cosh^{-2}{\Big(\frac{R^{i}}{\lambda_{D}}\Big)}\Bigg\{1 + \Big(\frac{R^{i}}{\lambda_{D}}\Big)\tanh{\Big(\frac{R^{i}}{\lambda_{D}}\Big)}\Bigg\}\Bigg].
\end{equation}
\end{widetext}
 
\noindent In the following, we will use Eqs.~(11) and (12) to evaluate the multiplicities of strangelets of the $i^{\rm th}$ species as defined in Eq.~(4). For doing this, we require an additional relation 

\begin{equation}
N^{i}_{u} = A^{i} + Z^{i},
\end{equation}

\noindent that is obtained from the definitions of the baryon number and the charge of a strangelet with $N^{i}_{u}$ being the number of the $u$-~quarks in that strangelet; see the discussion following Eq.~(3). In Eq.~(13), we have assumed that no electrons are localized inside strangelets. This  equation may be rewritten in the form of a transcendental equation, by using the definition $N^{i}_{u} = \Big(-\frac{\partial \Omega^{i}_{u}}{\partial \mu_{u}}\Big)_{T, {\sf{V}}^{i}}$ along with Eq.~(7) for $Z^{i}$, that involves the radius parameter $r^{i}_{\rm o}$ as defined in the discussion preceding Eq.~(5). This transcendental equation is solved iteratively to obtain the radius parameter of a particular species of strangelets corresponding to each trial value of the quark number chemical potential ($\mu_{q}$) of the strangelet-complex in equilibrium at freeze-out.

\section{Mass spectra of strangelets}

\noindent In this section, we explain the procedure adopted by us to numerically determine the multiplicities of various species of strangelet characterized by their baryon numbers or sizes in the strangelet-complex in thermodynamic equilibrium at freeze-out. To realize this aim, we first examine the condition of global charge-neutrality in the strangelet-complex. This condition is written as

\begin{equation}
\sum_{i} {\frac{\omega^{i}}{\cal V} Z^{i}} = \sum_{i} {n^{i}Z^{i}} = n_{e} \approx \frac{m_{s}^{6}}{192 \pi^{2}\mu_{q}^{3}}
\end{equation}

\noindent with $n^{i} = \frac{\omega^{i}}{\cal V}$ (which is the multiplicity density of strangelet-fragments of the $i^{\rm {th}}$ species). In deriving Eq.~(14), we have used the approximation for the electron chemical potential in the strangelet-complex that was adapted in the second paragraph in Sec.~II. Eq.~(4), along with Eqs.~(6)-(8) and (10)-(14), allow us to determine the values of $n^{i}$ for arbitrary positive integer values of $A^{i}$ after we self-consistently solve the above system of equations for the quark number chemical potential ($\mu_{q}$) in the strangelet-complex in thermodynamic equilibrium at freeze-out.

In the next step, we are required to determine the value of the available volume (${\cal V}$) of the strangelet-complex. This is done by using the condition for the conservation of the initial baryon number $A_{b}$~\cite{bib:bis12} of the strange matter released in an SS merger event. This condition is written in a form

\begin{equation}
{\cal V} = \frac{A_{b}}{\sum_{i}A^{i}n^{i}}.
\end{equation}

\noindent The value of the multiplicity $\omega^{i}$ of strangelets of the $i^{\rm th}$ species may now be easily determined from the known values of $n^{i}$ and ${\cal V}$. For the baryon number of the initial bulk matter, we choose $A_{b} = 1\times10^{53}$; this corresponds to a population averaged tidally released mass $M_{\rm {ejected}} \approx 10^{-4}$~$M_{\odot}$~\cite{bib:bauswein09} per binary SS merger obtained in the simulations with a bag value that corresponds to $B^{1/4} \approx 145$~MeV. This value of $B$ represents its lower bound determined by the fact that the energy per baryon of two-flavored quark matter must be higher than the one in $^{56}{\rm {Fe}}$~\cite{bib:kettner95, bib:mad99}, ie. $(E/A)_{u,d} \gtrsim 930$~MeV. Considering the limited accuracy of the MIT bag model, we may, as well, consider $B^{1/4} = 145$~MeV as the most favourable choice of the bag constant for which ordinary nuclei can decay into their strange quark phases only on a timescale longer than the age of the universe~\cite{bib:glend}. In Ref.~\cite{bib:bis12}, we considered this value of the bag constant to find the basic size distribution of strangelet-fragments; the standard bag value was taken to be $B^{1/4} = 145$~MeV as in Ref.~\cite{bib:haensel86}. In the model calculations of unpaired SQM, the bag value is, however, a bounded parameter that may be varied within the range $145~{\rm {MeV}} \leq B^{1/4} \leq 158~{\rm {MeV}}$ with its upper limit approximately corresponding to the limit of the absolute stability (ie. $E_{b}/A_{b} \lesssim 930~{\rm {MeV}}$) of bulk SQM at zero pressure and zero temperature~\cite{bib:haensel86, bib:gled92, bib:mad99, bib:bauswein09}. We have examined the consequence of the variation of bag value in this paper. 

It is important to note that the condition of global charge-neutrality (Eq.~(14)) was not relevant in Ref.~\cite{bib:bis12} in the case of fragmentation of the bulk SQM with vanishingly small quark masses. We, therefore, took the available volume (${\cal V}$) as a free parameter in that paper. There, we chose ${\cal V} = (2-9)V_{b} =(2-9)\times(\frac{4\pi}{3}r^{3}_{b}A_{b})$ by following the standard practice in the computations in nuclear fragmentation models~\cite{bib:randrup, bib:gross, bib:bondorf, bib:de07}; $V_{b}$ being the initial volume of the ejecta with $r_{b}$ as its bulk radius parameter. For an approximate estimate of $V_{b}$, we considered the value of the bulk radius parameter at zero temperature and zero pressure which is $r_{b} \approx (\frac{3}{4\pi n_{b}})^{1/3}$~\cite{bib:mad99}; ${n_{b}}$ being the baryon number density in that bulk SQM. Following Ref.~\cite{bib:mad99}, we considered $n_{b} = 0.7B^{3/4}$ (that corresponds to $m_{s} \rightarrow 0$ and $T = 0$) in Ref.~\cite{bib:bis12}. In this paper with massive $s$- quarks, the additional condition of global charge-neutrality (Eq.~(14)) in the strangelet-complex, along with the condition of the baryon number conservation (Eq.~(15)) in that complex, allow us to self-consistently determine an unique numerical value of ${\cal V}$. Here, we consider the approximation $n_{b} \approx \frac{1}{3}\Big [\frac{2 \mu_{b}^{3}}{\pi^{2}} + \frac{\mu_{b}^{3}}{\pi^{2}}(1 - \lambda^{2}_{sb})^{3/2}\Big]$~\cite{bib:mad99}, where $\lambda_{sb} = \frac{m_{s}}{\mu_{b}}$. The value of the quark number chemical potential ($\mu_{b}$) of the initial SQM ejecta is approximated as one third of the parameterized form of its energy ($E_{b}$) per baryon at $P^{i}_{\rm {ext}} = 0$ and $T = 0$, ie. $\mu_{b} = \frac{1}{3}(E_{b}/A_{b})$~\cite{bib:mad95, bib:mad99}, after the substitution of the appropriate value of $A_{b}$; the procedure to determine an approximate value of $E_{b}$ has been outlined in the discussion preceding Eq.~(4) in Sec.~II. An approximate value of the volume ($V_{b}$) of the initial bulk matter (with $m_{s} \ne 0$) may easily be determined by following the above prescription. With $A_{b} = 1\times10^{53}$, for example, the numerical value of the available volume turns out to be ${\cal V} \approx 4\times 10^{50}~{\rm {MeV}}^{-3} \approx 8\times 10^3V_{b}$. We also find that the available volume remains nearly the same for different bag values and for different values of the temperature (at freeze-out) lying within the corresponding ranges chosen in this paper. This available volume is, however, found to scale linearly with the value of the initial baryon number $A_{b}$.

Before we present the numerical results, we would like to add that, as in the case of nuclear disassembly models, the derived size distribution of strangelet fragments is sensitive to channel selection (ie. the selection of their baryon numbers). Some representative channels were selected in our earlier work in Ref.~\cite{bib:bis12}. In this paper, we instead consider all available positive integer values for $A^{i}$ of the fragment species to arrive at the number of fragments (ie. the multiplicity) pertaining to each species. While selecting those channels, we also take the charge numbers of strangelets into account. For this, we round off the real values obtained from Eq.~(7) to their nearest positive integers. The lower cutoff in the baryon number of a strangelet with $m_{s} = 95$~MeV is chosen so that the corresponding charge number becomes $Z^{i} = 1$ after rounding off.

\begin{figure}[htf!]
\subfigure[]
{\includegraphics[width=0.48\textwidth,clip,angle=0]{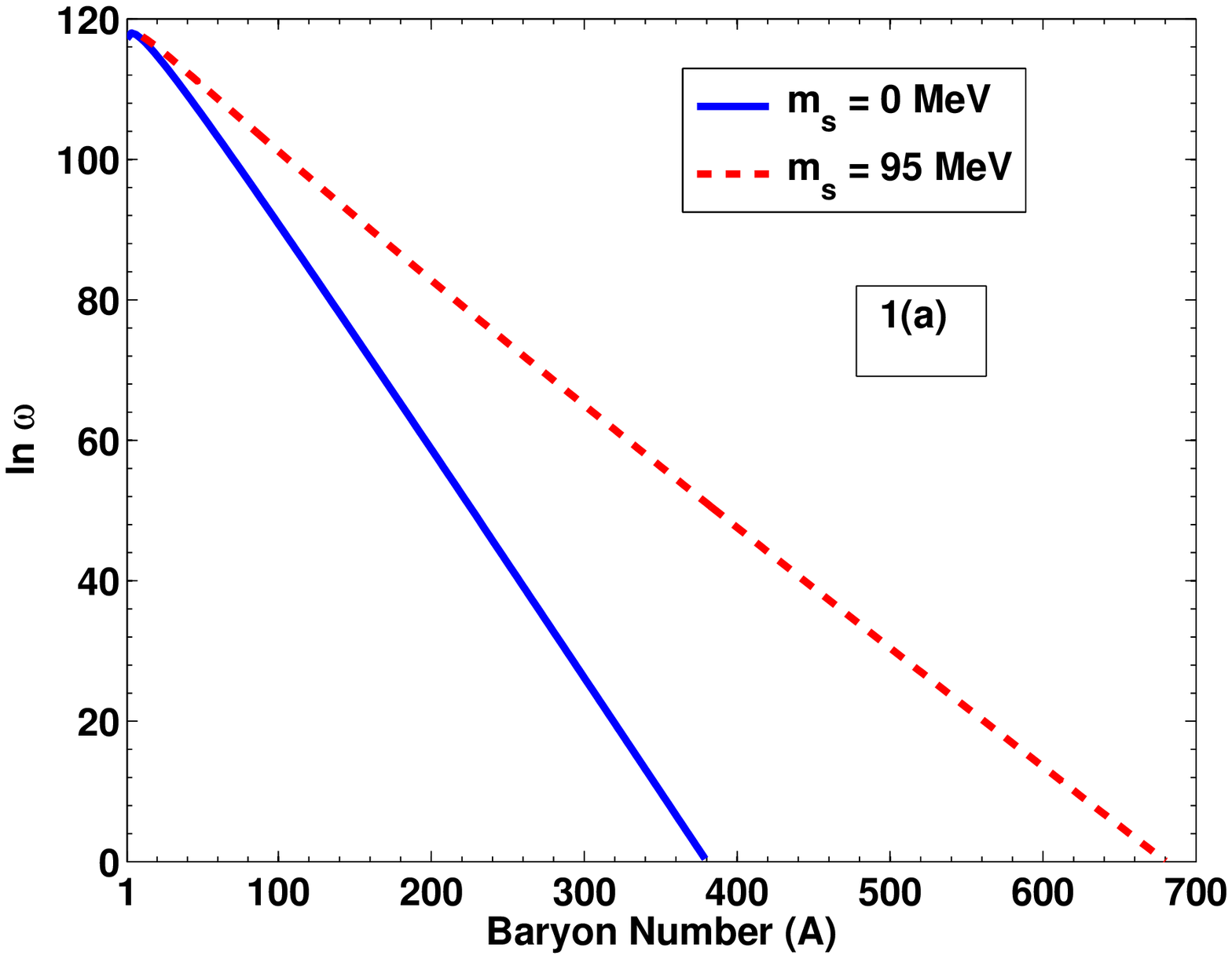}}
\subfigure[]
{\includegraphics[width=0.48\textwidth,clip,angle=0]{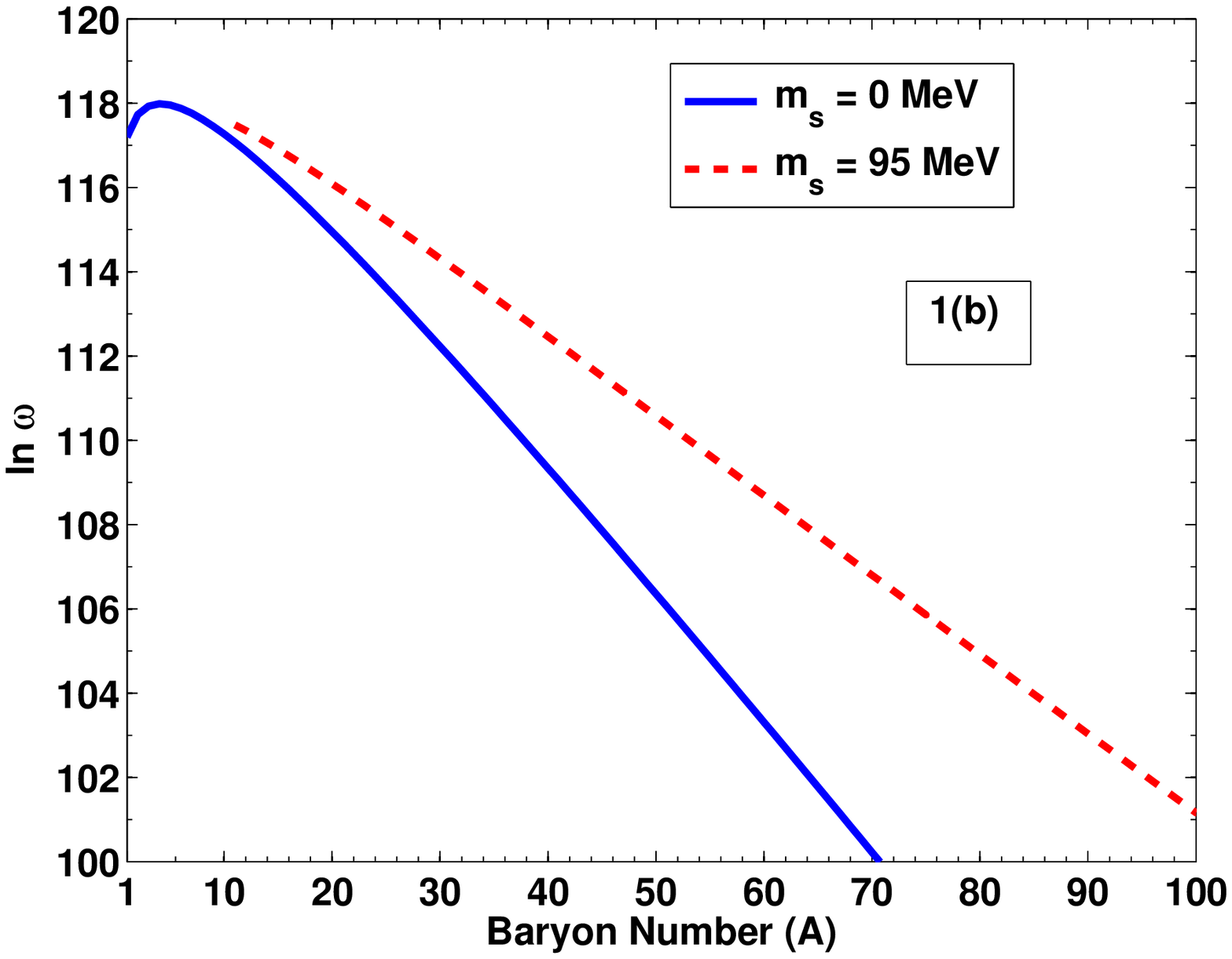}}
\caption{(color online) Multiplicity ($\ln\omega$) distribution of strangelets for massless ($m_{s}=0$) and massive ($m_{s}= 95$~MeV) $s$-~quarks both for a fixed value of the bag parameter ($B^{1/4} = 145~{\rm {MeV}}$) at a specific temperature ($T = 10~{\rm {keV}}$) at freeze-out. The $u$- and the $d$-~quarks are considered to be massless in both cases. The results are displayed for (a) the full range of the available baryon numbers and (b) for a limited range of baryon numbers of the strangelet-fragments. Fig.~1(b) is included to focus on the lower cutoff ($A \approx 11$ for $m_{s} = 95$~MeV) in the baryon numbers as well as the baryon number ($A \approx 4$) at which the peak of the distribution (for $m_{s} = 0$) is obtained.  Available volume is determined to be ${\cal V} \approx 8\times 10^{3}V_{b}$, $A_{\rm b} = 1\times 10^{53}$.}
\end{figure} 

Fig.~1(a) compares the multiplicities of strangelets in two cases, namely $m_{s} = 0$ and $m_{s} = 95$~MeV, for a fixed bag value ($B^{1/4} = 145~{\rm {MeV}}$) at a specific temperature ($T = 10~{\rm {keV}}$) at freeze-out. Fig.~1(b) displays the same in truncated baryon number range.  From these figures, it is apparent that the effect of $m_{s} \ne 0$ on the multiplicity distribution is not simply equivalent to an enhanced Boltzmann suppression as seems to have been recently suggested in Ref.~\cite{bib:paulucci14}. The distribution for $m_{s} = 0$ starts from $A^{i} \approx 1$, they are charge-neutral. This distribution has a peak at $A^{i} \approx 4$. A similar peak at $A^{i} \approx 4$ is seen for calculation with $m_{s} = 95~\rm{MeV}$, but since its charge is seen to be much less than one, we show the distribution from $A^{i} \approx 11$ which corresponds to $Z^{i} \approx 1$. The difference in the nature of the distribution for massless and massive s- quarks arises from a complex interplay of several factors.  Apart from giving rise to a lower cutoff at $A^{i}\approx 11$, incorporation of finite $m_{s}$ also leads to the suppression of lighter fragments and an enhanced production of heavier fragments. This is due to the quark mode surface tension that depends on finite mass of $s$-~quarks and vanishes in the limit of massless quarks according to the standard MIT bag model~\cite{bib:farhi,bib:berg1987,bib:mad99}. The surface term represents the energy required for creating the surface whereas the curvature term represents the energy required for bending it~\cite{bib:mardor}. As a consequence of the additional surface term, the total (surface $+$ curvature) requirement of energy for $m_{s} \ne 0$ is more than the energy required for curvature alone in the case $m_{s} = 0$. More energy is, therefore, required to produce small fragments out of the bulk SQM with massive $s$-~quarks. This has to be supplied from the limited reserve of thermal energy of the strangelet-complex at a fixed temperature. In statistical multifragmentation, an increase in the total (surface $+$ curvature) requirement of energy to form small strangelets (at a fixed temperature) results in a boost in the production of larger fragments at the cost of smaller fragments in a way such that the total baryon number is conserved. The converse leads to an enhanced production of lighter fragments at the cost of heavier fragments. These features of multifragmentation appear consistently in our results both in Ref.~\cite{bib:bis12} and in this paper. Such features of the disassembly model are independent of whether we consider massless or massive quarks as should become more apparent from the following discussions. Our preliminary calculations presented in Ref.~\cite{bib:bisposter} suggest that this nature of fragmentation is also independent of the choice of the CFL or the unpaired strangelets.      

\begin{figure}[htf]
\subfigure[]
{\includegraphics[width=0.48\textwidth,clip,angle=0]{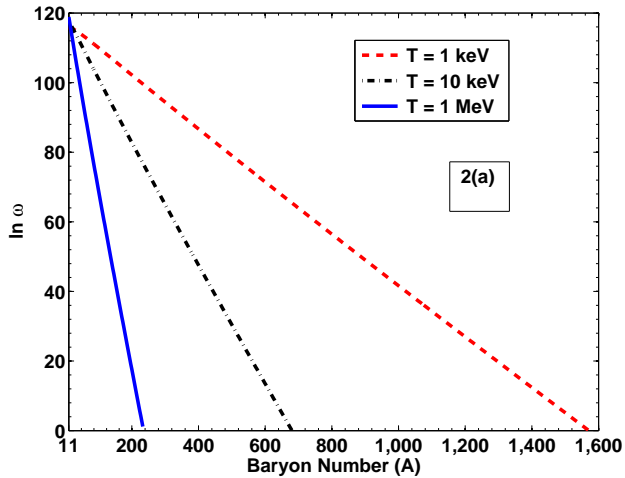}}
\subfigure[]
{\includegraphics[width=0.48\textwidth,clip,angle=0]{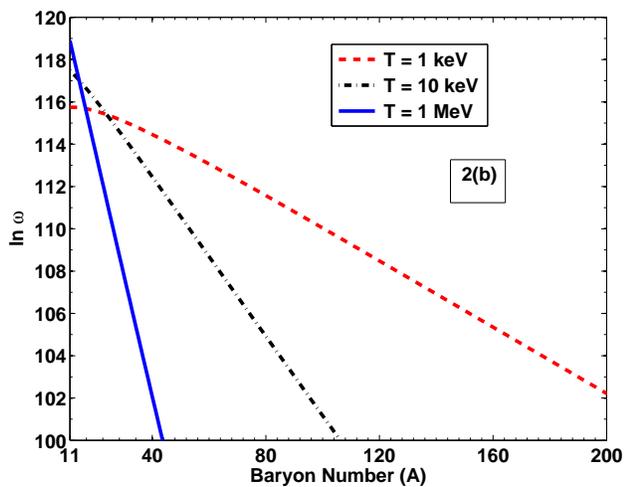}}
\caption{(color online) $\ln\omega$ vs. $A$ for the strangelet-fragments with $B^{1/4} = 145~{\rm {MeV}}$ and $m_{s} = 95~{\rm {MeV}}$ at three different temperatures at freeze-out. Variations are displayed for (a) the full range of possible baryon numbers and for (b) a limited range of baryon numbers of the fragments. Available volume is determined to be ${\cal V} \approx 8\times 10^{3}V_{b}$, $A_{\rm b} = 1\times 10^{53}$.}
\end{figure}

Figs.~2(a,b) display the size distributions of strangelet-fragments for a fixed bag value ($B^{1/4} = 145$~MeV) at three different temperatures, namely $T=1$~keV, $T=10$~keV and $T=1$~MeV, respectively. The variation of size distribution with changing temperature is in qualitative agreement with the one obtained in Ref.~\cite{bib:bis12} in the case of massless quarks. Suppression of heavier fragments and enhanced production of lighter fragments with increasing temperature are noted for $m_{s} = 0$~\cite{bib:bis12} and also for $m_{s} \ne 0$. The distributions are separately plotted in Fig.~2(b) for a limited range of baryon numbers for the sake of clarity. The progressive shift of the distribution to lower masses with increasing temperature may be noted. Such results are commonplace in the case of nuclear fragmentation~\cite{bib:gross, bib:bondorf, bib:de07}. 

It is known that the surface free energies and the curvature energies of both the baryonic (ie.~nuclei) and the quasi-baryonic (ie.~SQM) fragments  decrease with increasing temperature~\cite{bib:de12, bib:paulucci08}. This, in turn, implies that the total requirement of (surface $+$ curvature) energy to produce small strangelets out of the initial bulk matter is reduced at higher temperature. Such reduced requirement of energy is easily met by a larger reserve of thermal energy of the strangelet-complex at an enhanced temperature. This, along with the condition for the conservation of baryon number, ensure copious production of lighter fragments and suppressed production of heavier fragments with increasing temperature. Such pattern of decreasing fragment-sizes with increasing temperature is in consonance with the standard results of nuclear fragmentation models~\cite{bib:gross, bib:de07, bib:rosswog}. Recent discussion on fragmentation in Ref.~\cite{bib:paulucci14} finds an opposite tendency in the variation of size distribution of CFL strangelets with changing temperature. The authors of Ref.~\cite{bib:paulucci14} seem to attribute this behavior of the fragmentation derived by them to the finite mass of $s$-~quarks combined with the color-superconductivity of the strangelets. It is relevant here to add that an earlier exploratory work~\cite{bib:bisposter} of ours found that the changes in the frequency distribution of CFL strangelets, having massless quarks, with changing temperature are in qualitative agreement with Figs.~2(a,b).    

\begin{figure}[htf]
{\includegraphics[width=0.48\textwidth,clip,angle=0]{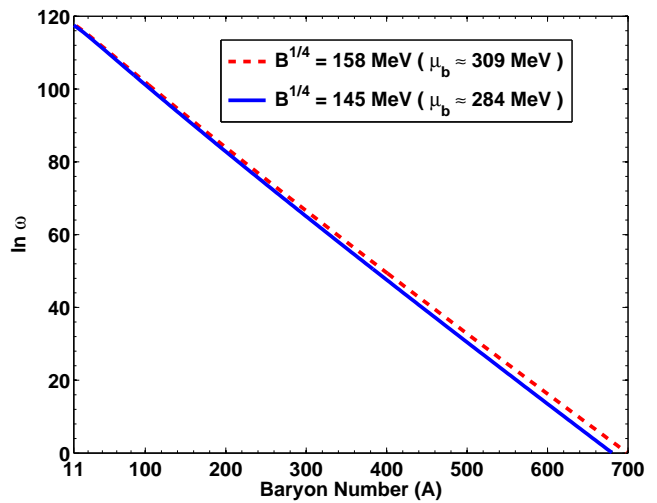}}
\caption{(color online) $\ln \omega$ vs. $A$ for strangelets with $m_{s} = 95~{\rm {MeV}}$ at a specific temperature ($T=10$~keV) but at two different bag values as indicated in the diagram. Approximate value of the quark number chemical potential ($\mu_{b}$) of the initial bulk matter before fragmentation, that corresponds to each value of the bag parameter at zero external pressure and zero temperature, is also displayed. Available volume is determined to be ${\cal V} \approx 8\times 10^{3}V_{b}$, $A_{\rm b} = 1\times 10^{53}$.}
\end{figure}

In Fig.~3, we examine the influence of bag values on the fragmentation pattern of strangelets for $m_{s} = 95$~MeV at a fixed temperature ($T = 10~{\rm {keV}}$) at freeze-out. With other parameters remaining the same, an enhanced bag value increases the quark number chemical potential in the strangelet-complex that, in turn, increases the surface and curvature energies of the strangelet-fragments. This obviously enhances the energy requirement for the formation of light fragments, as a consequence of which the production of heavier fragments at the cost of the lighter ones is preferred. Small variation of the fragmentation pattern with the variation of bag value in Fig.~3 may also be interpreted in terms of an increase in the quark number chemical potential ($\mu_{b}$) of the initial bulk SQM due to an increase in the value of the bag parameter $B$. The resulting increase in the baryon number density of the initial bulk matter~\cite{bib:mad99} would favour larger fragments in agreement with the standard results of the nuclear fragmentation models~\cite{bib:rosswog}. In Fig.~3, the lower cutoff (corresponding to $Z^{i} \approx 1$) in the baryon number  of the distribution changes from $A^{i} \approx 11$ for $B^{1/4} = 145$~MeV (that corresponds to $\mu_{b} \approx 284~{\rm {MeV}}$) to $A^{i} \approx 14$ for $B^{1/4} = 158$~MeV (that corresponds to $\mu_{b} \approx 309~{\rm {MeV}}$). We recall that the procedure to determine an approximate value of the quark number chemical potential $\mu_{b}$ of the initially ejected bulk matter determined at zero external pressure and zero temperature was pointed out in the third paragraph of this section.

The comparison displayed in Fig.~3 is a convenient way of demonstrating the effect of $B$ on the fragmentation pattern in which the baryon number ($A_{b}$) of the initial bulk matter is taken to be the same for both the bag values. Preliminary simulations~\cite{bib:bauswein09} of SS merger, however, find no mass ejection in the case $B^{1/4} \approx 158$~MeV due to the resulting compactness of the merging SSs. These simulations seem to indicate that, for $B^{1/4} \sim 158$~MeV, the merger product collapses into a black hole (BH) faster than the time required for the formation of its tidal arms. Although the actual simulations in Ref.~\cite{bib:bauswein09} were done only at two nearly extreme bag values in the range $145~{\rm {MeV}} \lesssim B^{1/4} \lesssim 158~{\rm {MeV}}$, the authors of that work expect that the population averaged ejecta mass ($M_{\rm {ejected}}$) for any intermediate bag value within the above interval would lie somewhere in the range $10^{-4} M_{\odot} > M_{\rm {ejected}} \gtrsim 0$ with lesser amount of ejected mass corresponding to a larger bag value. We have checked that the shape of the fragmentation pattern corresponding to a particular bag value remains almost invariant for any reduced value of the mass of the initially released bulk matter except that all the multiplicities are now reduced by an appropriate factor from the ones obtained for $M_{\rm ejected} = 10^{-4} M_{\odot}$ (ie. $A_{b} = 1\times 10^{53}$). Such scaling makes it convenient to estimate the possible fluxes of strangelets in PCR that correspond to various mass distributions of strangelets injected in the Galaxy for different bag values.

\section{Stability of the produced fragments}

\begin{figure}[htf]
{\includegraphics[width=0.48\textwidth,clip,angle=0]{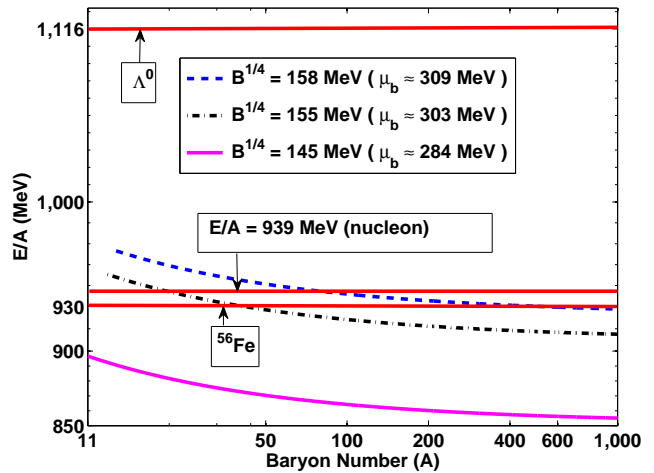}}
\caption{(color online) Variation of the energy per baryon ($E/A$) against changing baryon number ($A$) of the strangelet-fragments (with $m_{s} = 95~{\rm {MeV}}$) for three different values of the bag parameter at a specific temperature ($T=1$~MeV) at freeze-out. Corresponding value of the quark number chemical potential ($\mu_{b}$) of the initial bulk matter at zero pressure ($P_{\rm {ext}}=0$) and zero temperature is displayed against each bag value for the sake of comparison. The solid (red) horizontal lines mark the energies per baryon of $^{56}{\rm Fe}$, nucleon and  $\Lambda^{0}$-hyperon, respectively, that delineate the thresholds for absolute stability, metastability and instability of the fragments. Available volume is determined to be ${\cal V} \approx 8\times 10^{3}V_{b}$, $A_{\rm b} = 1\times 10^{53}$.}
\end{figure}

\noindent Having determined the size distribution of the strangelet-fragments, that may be  injected in the Galaxy by SS merger, we investigate which ones of these fragments are stable with respect to $^{56}{\rm {Fe}}$ nucleus; their energy per baryon ($E^{i}/A^{i}$) should be less than $930$~MeV. Those strangelets represent the true ground state of hadronic matter, the detection of which would make for an important discovery. Apart from this novelty, those strangelets are possibly the only ones to survive during the plausible confinement-time ($\sim 10^{7}~{\rm {yr}}$~\cite{bib:mad2002}) of strangelets in the Galaxy. Such strangelets are easily detectable in PCR in the solar neighborhood. It was, however, pointed out in Ref.~\cite{bib:glend} that all possible values of the model parameters, that place the energy per baryon of  strangelets in the vicinity of that of nuclear matter cannot be discarded. In fact, precise values of $E^{i}/A^{i}$ of strangelets, ie. whether they lie marginally above the nucleon mass or below the energy per nucleon in $^{56}{\rm {Fe}}$, is a matter that involves only $\sim 1\%$ deviation in numerical calculations. A deviation of this magnitude may be insignificant in view of the uncertainties in the accuracy of the results derived from the MIT bag model~\cite{bib:glend}. Keeping this issue in mind, we examine the values of $E^{i}/A^{i}$ of strangelet fragments as a function of their baryon number $A^{i}$ for three different values of the bag parameter at a temperature $T = 1$~MeV at freeze-out; the stability of strangelets is known to increase at lower temperatures~\cite{bib:he}. The results of this investigation are displayed in Fig.~4. As in Fig.~3, we indicate the value of the quark number chemical potential ($\mu_{b}$) of bulk matter at zero pressure and zero temperature against each bag value in this figure also. In Fig.~4, the solid (red) horizontal lines mark the values of the energy per baryon of $^{56}{\rm Fe}~(E/A = 930~{\rm {MeV}})$ nucleus, nucleons  ($E/A = 939~{\rm {MeV}}$) and $\Lambda^{0}$-hyperons ($E/A = 1116~{\rm {MeV}}$), respectively. According to Ref.~\cite{bib:glend}, the latter two lines delineate the thresholds for stability ($E^{i}/A^{i} \lesssim 939~{\rm {MeV}}$), metastability ($939~{\rm {MeV}} < E^{i}/A^{i} < 1116~{\rm {MeV}}$) and instability ($E^{i}/A^{i} \gtrsim 1116~{\rm {MeV}}$) of the strangelet-fragments; see Refs.~\cite{bib:schaff1997} for details. In this paper, we avoid a detailed examination of the life-times and the decay modes of metastable and unstable strangelets. Instead, we simply look for fragment-sizes satisfying the stability criterion in Ref.~\cite{bib:glend} for each value of $B^{1/4}~({\rm {or}}~\mu_{b})$ displayed in Fig.~4. In this figure, we find that all the strangelets having $A^{i} \gtrsim 11$ are stable relative to the $^{56}{\rm Fe}$ nucleus for $B^{1/4} = 145~{\rm {MeV}}$~(ie.~$\mu_{b} \approx 284~{\rm {MeV}}$). For $B^{1/4} = 155~{\rm {MeV}}$~(ie.~$\mu_{b} \approx 303~{\rm {MeV}}$) and $B^{1/4} = 158$~MeV~(ie.~$\mu_{b} \approx 309$~MeV), the strangelets having their sizes in the respective ranges $A^{i} \gtrsim 23$ and $A^{i} \gtrsim 90$ are stable relative to the nucleons. 

Here, it is relevant to take note of an altogether different scenario of fragmentation of a (positively charged) strangelet (embedded in a charge-neutralizing cloud of electrons), with its radius satisfying $R \gg \lambda_{D}$, through the ``fission instability"  proposed in Refs.~\cite{bib:alford, bib:alford08, bib:alford2012, bib:jaikumar2006} within a model-independent theoretical framework. This instability affects even the cold ($T = 0$) strangelets. The onset of this instability depends crucially on the surface tension ($\sigma_{s}$) at the boundary of the quark matter. Instability sets in whenever $\sigma_{s} < \sigma_{\rm {crit}}$, $\sigma_{\rm {crit}}$ being a critical surface tension whose values have been determined in Ref.~\cite{bib:alford} in the case of the MIT bag model for a wide range of values of $m_{s}$ along with different values of the quark number chemical potential ($\mu_{b}$) of the absolutely stable, charge-neutral bulk SQM at zero external pressure and zero temperature. For any particular value of $m_{s}$, the parameter $\mu_{b}$ represents the bag value as it did in Figs.~3 and 4. For $\mu_{b} = 305$~MeV, for example, the critical surface tension takes on values in the range $0.1~{\rm {MeV}}~{\rm {fm}}^{-2} \le \sigma_{\rm {crit}} \lesssim 2.7$~MeV~${\rm {fm}}^{-2}$ for the mass of the $s$-~quarks lying in the range $100~{\rm {MeV}} \le m_{s} \le 240~{\rm {MeV}}$; the upper bound corresponding to the strangelets at the threshold of their absolute stability, ie. $\mu_{q} \lesssim 310~{\rm {MeV}}$~\cite{bib:alford}. Here, we wish to point out that, as a consequence of not taking the effects of dynamical or explicit chiral symmetry breakdown into account (ie. without any source of the constituent quark-masses~\cite{bib:hartmann1999}), the mass ($m_{s}$) that enters in the traditional MIT bag model for the unpaired, noninteracting SQM (as in Ref.~\cite{bib:alford}) can only be the current mass of the $s$-~quarks. As we discussed in Sec.~I, the value of this mass has recently been estimated to be $m_{s} \lesssim 100$~MeV with reasonable accuracy. In view of this development, the possible value of the critical surface tension in the MIT bag model (estimated from Fig.~3 in Ref.~\cite{bib:alford}) seems to be $\sigma_{\rm {crit}}\sim 0.1$~MeV~${\rm {fm}}^{-2}$ for values of $\mu_{b}$ lying in the range $(284-309)$~MeV; see Fig.~4 for the corresponding range of approximate bag values. Such value of $\sigma_{\rm {crit}}$ is at least an order of magnitude smaller than the typical values of the quark mode surface tension ($\sigma_{s} \sim (5-10)$~MeV~${\rm {fm}}^{-2}$;  see Sec.~I) in the MIT bag model of strangelets. Above comparison seems to suggest that the stable strangelets, with their sizes in the range $R^{i} \sim ($0.4-2.2$)\lambda_{D}$ that we obtain in this paper, are also stable against the fission instability proposed in Refs.~\cite{bib:jaikumar2006, bib:alford, bib:alford08, bib:alford2012}. In the next section, our aim would be to find an order of magnitude estimate of the integrated (over baryon numbers) intensity of those strangelet-fragments in the vicinity of the solar system.

\section{Discussion}

\noindent Apart from conventional NSs, SMH predicts the existence of a new family of compact stars, namely, the SSs~\cite{bib:itoh, bib:witten, bib:haensel86}. These SSs result from the decay of metastable NSs into more bound configurations \textit{via}~different possible routes; see Refs.~\cite{bib:angeles2010} and the citations therein. Here, we confine our attention to the debris of possible collisions between SSs that may be a major source of strangelets in PCR~\cite{bib:mad99, bib:mad2002, bib:mad2005, bib:bauswein09}. Earlier~\cite{bib:bis12}, we attempted to estimate the intensity of those strangelets integrated over baryon numbers in PCR in the limit $m_{s} \rightarrow 0$ by employing a diffusion approximation found from Ref.~\cite{bib:ginz}. In this paper, we improve upon that estimate by incorporating the effects of finite $m_{s}$ as well as a wider range of permissible $B$ values. Assuming a rate $\sim 10^{-5}~{\rm yr}^{-1}$~\cite{bib:bauswein09, bib:belczynski} of SS merger in each Galaxy and assuming the resulting strangelets to spread homogeneously in a galactic halo of radius $\sim 10$~kpc~\cite{bib:mad99} within their galactic confinement time, an approximate intensity of strangelets of the $i^{\rm {th}}$ species in the solar neighborhood was written in Ref.~\cite{bib:bis12} as

\begin{equation}
I(A^{i}) \sim 5\times10^{-48}\omega^{i}~{\rm {particles}}~{\rm m}^{-2}~{\rm {sr}}^{-1}{\rm {yr}}^{-1}.
\end{equation}

\noindent Here, $\omega^{i}$ is the multiplicity of strangelets of the $i^{\rm {th}}$ species as defined in Eq.~(4) which was, in turn, derived from the standard formula of SMM (Eq.~(3)) in Ref.~\cite{bib:bondorf}; the quantities $\cal V$ and ${\cal L}^{i}$ in Eq.~(3) were defined in Sec.~II. Numerical values of $\omega^{i}$ have been determined in Sec.~IV from the thermodynamic properties of strangelets after satisfying the conditions of charge-neutrality and the baryon number conservation in the strangelet-complex.

\begin{table*}[htf]
\caption{Expected ranges of the integrated (over baryon number) intensity of stable, unpaired strangelets in the solar neighborhood for different intervals of plausible bag values and for the corresponding ranges of the (tentatively) estimated tidally released mass per SS merger. The estimations of ejected masses are inspired by the recent simulations~\cite{bib:bauswein09} in which the limit of mass-resolution was $\sim 10^{-5} M_{\odot}$.}
\begin{center}

  \begin{tabular}{ |p{2cm}| p {2cm}| p{4cm} |p{6cm}| }
    \hline
    \hline
    $B^{1/4}$ (MeV) &  $\mu_{b}$~(MeV) & Mass of strange matter released per SS merger ($M_{\odot}$) & Estimated integrated intensity of stable strangelets ($\rm{particles~m^{-2}}~\rm{sr^{-1}}~\rm{yr^{-1}}$) \\
\hline
    $145$ & $\approx 284$ & $ \sim 10^{-4}$& $\sim (2-5)\times 10^{4}$ \\ 
    $(146-150)$ & $\approx (286-294)$ & $ \sim (0.01-1.0) \times 10^{-4}$ & $\sim (2-500) \times 10^{2}$ \\
    $(151 -158)$ & $\approx (296-309)$ &$ \sim (0.0-1.0) \times 10^{-6}$ & $\sim (0-2) \times 10^{2}$ \\
    \hline
     \hline
    \end{tabular}
\end{center}
\end{table*}

Approximation~(16) provides only an order of magnitude estimate. Important issue of the acceleration of the strangelets by the astrophysical shock waves has been left out of consideration. In this estimate, the diffusion coefficient of galactic strangelets is not determined from rigorous calculations. It also ignores the possible interaction of strangelets with the interstellar medium. Moreover, it does not take the effects of the geomagnetic field and the solar modulation into consideration. In the particular case $B^{1/4} \approx 145$~MeV, simulations find a population averaged tidally released mass $M_{\rm {ejected}} \approx 10^{-4}~M_{\odot}$ per SS merger. A summation of the estimate~(16) over the values of $\omega^{i}$ for all the stable ($A^{i} \gtrsim 11$) fragments, obtained from the results displayed in Figs.~2 and 4, yields the integrated strangelet intensity in the the solar neighborhood in this situation. The values of this intensity lie within the range $\sim (2-5)\times 10^{4}~{\rm {particles}}~{\rm m}^{-2}~{\rm {sr}}^{-1}{\rm {yr}}^{-1}$ depending on the formation temperature of the strangelets. Increasing the value of $B$ within the range $145~{\rm {MeV}}< B^{1/4} \lesssim 158~{\rm {MeV}}$ has an appreciable effect. In that case, we are required to reduce the average tidally released mass per SS merger to values within the corresponding range $10^{-4}~M_{\odot}> M_{\rm {ejected}} \gtrsim 0$ to comply with the results of the recent simulations. However, those simulations were performed only at two bag values near the upper and the lower ends of the aforesaid interval. Precise value of tidally released mass for an intermediate $B$ cannot  be determined from those simulations. Such uncertainty notwithstanding, in Table 1, we display the estimated ranges of integrated strangelet flux in the solar neighborhood for different intervals of bag values. In this table, we also display the tentative ranges of values of the average mass ($M_{\rm {ejected}}$) released per SS merger for different intervals of bag values with the {\textit{caveat}} that the amounts quoted in Table 1 for intermediate $B$ values are presented only for the sake of an illustration. The actual amount of this mass for an intermediate bag value can only be determined through detailed high resolution simulations of SS merger for a number of bag values lying within the range $145~{\rm {MeV}} < B^{1/4} < 158~{\rm {MeV}}$. Such detailed simulations are yet to be performed. In Table 1, large dispersions in the estimated fluxes for different bag values reflect on such limitation of the recent simulations in scanning the parameter space. They also reflect on the limited mass-resolution of the present simulations and also on the theoretical uncertainty in predicting the formation temperature of the strangelets. In the case $B^{1/4} \sim 158$~MeV, for example, the simulated results predict a vanishing strangelet flux in the solar neighborhood. For an assumed tidally ejected mass $M_{\rm {ejected}} \sim 10^{-6} M_{\odot}$ per stellar merger (that is an order of magnitude smaller than the limit of mass-resolution of the existing simulations), the approximation (16), on the other hand, yields an integrated flux $\sim 1~{\rm {particle}}~{\rm m}^{-2}~{\rm {sr}}^{-1}{\rm {yr}}^{-1}$ at a sufficiently low temperature ($T \sim 1~{\rm {keV}}$) at freeze-out in this particular case. Such flux is, in principle, measurable in the observations with the detector systems being similar to the one installed in AMS-02 experiment at the present level of its sensitivity~\cite{bib:kounine}.

Table 1 predicts measurable fluxes of stable, ordinary strangelets in PCR for a reasonably wide range (i.e. $145~{\rm {MeV}} \lesssim B^{1/4} \lesssim 158~{\rm {MeV}}$) of bag values. The results displayed in this table seem to disapprove of the claim in Ref.~\cite{bib:paulucci14} that the existence of a large number of (stable) strangelets in cosmic rays is highly unlikely and would certainly be negligible if color superconductivity is not considered. Of course, nobody can deny the  importance of studying multifragmentation of CFL matter leading to some sort of size distribution of CFL strangelets. We also note that such a study has already been undertaken in Ref.~\cite{bib:paulucci14} by adapting a nuclear liquid-gas phase transition model~\cite{bib:bug}. In carrying out this calculation, the authors of Ref.~\cite{bib:paulucci14} have found certain ambiguity (or inconsistency) in their results that has led them to conclude that either most of the CFL matter does not fragment at all or the standard techniques of SMM are inadequate to describe the fragmentation of CFL SQM. In this context, we would like to point out that our preliminary results~\cite{bib:bisposter} on multifragmentation of CFL matter in the limit of massless quarks was free from such inconsistency.

The authors of Ref.~\cite{bib:paulucci14} have assumed the CFL strangelets to be more abundant in PCR because of their greater stability in comparison with the ordinary ones. Accepting the importance of the detection of CFL strangelets in PCR, we still have some doubt regarding  the feasibility of a mechanism producing those strangelets. Recent hydrodynamical simulations~\cite{bib:herzog, bib:paulucci14} suggest that no CFL SQM is likely to be ejected outside the surface of the NS during the conversion of its interior into CFL matter. The combustion front would stop before it reaches the stellar surface. On the other hand, the present scenario of tidally released quark matter in SS merger cannot be extended straightway to CFL strangelets. This is due to the recent arguments~\cite{bib:mad00} against the possibility of the observed cold compact stars being bare CFL stars (CFLSs). To circumvent this problem, Ouyed \textit{et al.}~\cite{bib:ouyed09} have invoked a scenario of collisions between hot and young CFLSs and their NS companions in compact binary stellar systems of the Galaxy that may produce CFL strangelets. These authors find an estimate $\sim (1-100)\times 10^{2}~{\rm {particles}}~{\rm m}^{-2}~{\rm {sr}}^{-1}{\rm {yr}}^{-1}$ for the integrated flux of CFL strangelets in the solar neighborhood. A detailed derivation of that estimate is, however, unavailable in Ref.~\cite{bib:ouyed09}.
Although the present paper is focused on the availability of unpaired strangelets in PCR, we may still like to use approximation~(16), along with the determined size distribution of strangelets in Sec.~IV, to get a rough idea regarding the order of magnitude of the possible intensity of CFL strangelets \textit{vis-a-vis} the unpaired ones in the vicinity of the solar system. For this purpose, we first note that, for a ``not unreasonable" value of the pairing energy gap $\Delta = 100$~MeV~\cite{bib:raja01}, the lower bound of bag values for the stable CFL matter at zero temperature is taken as $B^{1/4} \gtrsim 156$~MeV to avoid spontaneous decay of an ordinary nucleus into a two-flavor color superconducting phase~\cite{bib:mad01}. Accordingly, a simple extrapolation of approximation (16) to the case $B^{1/4} \sim 156$~MeV after the substitution of the possible lower bound $\sim 10^{-7}~{\rm {yr}}^{-1}$~\cite{bib:ouyed09} of the rate of CFLS-NS collisions in the Galaxy along with an assumed tidally released CFL mass $\sim 10^{-6}~M_{\odot}$ in each of such collisions is likely to bring down the integrated flux of CFL strangelets somewhere within the range $\sim (10^{-4}-1.0)~{\rm {particles}}~{\rm m}^{-2}~{\rm {sr}}^{-1}{\rm {yr}}^{-1}$. Further improvement of such estimate would require the determination of actual size distribution of CFL strangelets which would be the subject matter of a separate paper. We however note that, an extrapolation of the results of the recent simulations on SS merger seems to indicate that, due to the supposedly compact nature of the CFLSs resulting from their larger binding energy, there is a possibility that the entire product of the CFLS-NS merger may collapse into a Black hole before the tidal forces have sufficient time to spew appreciable CFL mass out of the gravitational influence of the combined system. We could hardly expect to detect any CFL strangelet in PCR in that case.

The ultimate vindication of SMH would depend on the detection of either unpaired or CFL or both the types of strangelets in PCR. In this paper, we have examined a plausible model of the rate of injection of unpaired strangelets in the Galaxy. A separate study of the possible mass distribution of CFL strangelets at their source as well as an examination of the more sophisticated galactic propagation models for both the types of strangelets are required to arrive at a definite prediction of strangelet-flux in PCR for AMS-02 and other potential experiments.


\end{document}